\documentclass[11pt,leqno]{article}%
\usepackage{amssymb,bbm,amsmath,
amsthm,
amscd}
\usepackage{vmargin}
\usepackage[dvipsnames]{xcolor}
\usepackage{
 array}
 
 \usepackage{graphicx}
\catcode`\@=11
\newbox\bz@
\newdimen\bdimz@
\def\linethrough#1{\setbox\bz@=\hbox{#1}%
\bdimz@=\ht\bz@ \divide\bdimz@ by 5 \advance\bdimz@ by -\dp\bz@ \ht\bz@=\bdimz@
\leavevmode\hbox{$\overline{\overline{\box\bz@}}$\relax}}
\catcode`\@=12

\def\nbZ{{\mathchoice {\hbox{$\sf\textstyle Z\kern-0.4em Z$}}
{\hbox{$\sf\textstyle Z\kern-0.4em Z$}}
{\hbox{$\sf\scriptstyle Z\kern-0.3em Z$}}
{\hbox{$\sf\scriptscriptstyle Z\kern-0.2em Z$}}}}

\newcounter{compteur}[subsection]

\newcommand{\step}{ \refstepcounter{compteur} {\arabic{section}.\arabic{subsection}.\roman{compteur}}.  
}

\usepackage[colorlinks=true]{hyperref}
\newtheorem{theorem}{Theorem}[section]
\newtheorem{lemma}[theorem]{Lemma}

\usepackage{macros}

\makeatletter

\@addtoreset{equation}{section}

\begin{document}

\author{Tom\'as Chac\'on Rebollo \thanks{
Dpto.\ EDAN \& IMUS, Universidad de Sevilla, C/Tarfia, s/n. 41012~Sevilla, Spain \& BCAM-Basque Center for Applied Mathematics, Avda. Mazarredo, 14. 48009 Bilbao, Spain; {\tt chacon@us.es}
}, Roger Lewandowski\thanks{Mathematical Research Institute of Rennes, UMR CNRS 6625, European University of Brittany  \& Fluminance team, INRIA, Bat 22, Campus Beaulieu, 35042 Rennes cedex, France; {\tt Roger.Lewandowski@me.com}}  }
\title{A Variational Finite Element Model for Large-Eddy Simulations of Turbulent Flows}

\maketitle
\begin{abstract} We introduce a new Large Eddy Simulation model  in a channel, based on the projection on finite element spaces as filtering operation in its variational form, for a given triangulation $\{ {\cal T}_h \}_{h>0}$. The eddy viscosity is expressed in terms of the friction velocity  in the boundary layer due to the wall, and is of a standard sub grid-model  form outside the boundary layer. The mixing length scale is locally equal to the grid size. The computational domain is the channel without the linear sub-layer of the boundary layer.  The no slip boundary condition (BC) is replaced by a Navier (BC) at the computational wall. 
Considering the steady state case, we show that the variational finite element model we have introduced, has a solution $(\vv_h, p_h)_{h>0}$ that converges to a solution of the steady state Navier-Stokes Equation with Navier BC.  
 \end{abstract}

 MCS Classification : 76D05, 76F65, 65M60,

\section{Introduction}\label{Chapter_Numeric_SGM_steady}

Numerical simulations  of incompressible turbulent flows cannot be performed from  the evolutionary Navier-Stokes Equations (NSE), 
\BEQ \begin{array}{rcll}
\p_t \vv + (\vv \cdot \g) \vv-\nu \Delta \vv+\nabla p&=&\fv, \\
\div \vv &=&0, 
\end{array} 
\EEQ
because of a great computational complexity due to the structure of the turbulence \cite{SP00}. This is why various mathematical models derived from the NSE are used to simulate some features of turbulent flows, such as their statistical means or their large scales motions, this last way being known as "Large-Eddy Simulation" (LES), which is our concern in the present paper. 
\medskip

LES has attracted much attention these  last two decades, especially because of the increasing of computational ressources, enabling to enlarge the range of scales  that LES models might simulate. Basically, LES aims at computing filtered fields such as $\overline \vv = G \star \vv$, $G$ being a smooth transfer function \cite{LMC2005, SP00, RFBP1997, sagaut}. The filtering operation also might be carried out by solving PDE's \cite{BL12,  layton, germ1, FHT01, LL02, LL03}. 
\medskip

Stresses that appear by  filtering the non linear term $(\vv \cdot \g) \vv$ in the NSE, are considered to be diffusive, therefore often modeled by a turbulent diffusion term such as $- \div (\nut \g \overline \vv)$, where $\nut$ is an eddy viscosity. One challenge of the LES is the determination of $\nut$. 
\medskip

In this paper, we study the case of a channel flow, periodic in the $x_1-x_2$ axis for simplicity. 
The first  idea is that the projection on finite element spaces, based on a given triangulation $\{ {\cal T}_h \}_{h>0}$, is a natural filtering operation, so that we seek for $\vv_h$ instead of $\overline \vv$, where $\vv_h$ is the projection of $\vv$ on a suitable finite element space $\vbh$. The second idea is that one can specifically model the eddy viscosity on the boundary layer by means of wall laws.  
\medskip

Indeed, following Kolmogorov theory \cite{AK41a}, we consider the turbulence to be isotropic at scales small enough inside the flow domain. 
This assumption yields to take the eddy viscosity of a Kolmogorov-Prandtl-Smagorinsky form,  $\nut = h^2 | D \vv_h |$\footnote{ 
$D \vv_h = (1/2) (\g \vv_h + \g \vv_h^t)$}, $h$ being the mixing length, which is the standard 
sub-grid model (SGM) \cite{chalew}.   
\medskip

However, near the wall, turbulence is not isotropic and complexity is higher than far from the wall \cite{PB02}, so that standard SGM cannot be used there. Usual methods proceed as follows. 
\medskip

First one uses the known structure of the turbulent boundary layer, as initially described by von  K\'arm\'an \cite{TK30} and fully developed by Schlichting \cite{schli}.  Basically, the boundary layer may be split into two sub-layers, the linear sub-layer where 
the mean velocity profile is linear, and next, the log sub-layer where the mean velocity profile is specified by a log function.
Notice that one can consider more sophisticated models to model the boundary layer \cite{spald}, nevertheless always  involving a log law. In all cases, those models involve an essential quantity which is the friction velocity $u_\star$ (see (\ref{eq:friction}) in  \ref{step:friction1} below). 

\medskip

Next, one splits the domain into two subdomains, the boundary layer, and the computational domain which is the domain's part  not containing 
the boundary layer. One then uses non linear boundary conditions at boundaries of the computational domain such as wall laws \cite{RL97, MP94, pares}. 
\medskip

Based on the fact that today more computational resources are available to increase accuracy for simulating the mean flow inside 
the log layer, we take as computational domain the domain's part  without 
the linear sub -layer,
using  an eddy viscosity  of the form $\nut = h u_\star$ inside de log layer, deduced from standard dimensional analysis \cite{chalew, RL97}. 

\medskip

To conclude the modeling process, it remains to: i)  specify how $u_\star$ is calculated, ii) specify boundary conditions (BC) at  computational domain boundaries, iii) fix the choice of the mixing length scale. 
\medskip

i) We assume that log law holds inside the boundary layer. Thanks to invertibility of the non linear profil, we can define $u_\star$ as 
$u_\star (\vv, \x)$, that satisfies suitable estimates (see \ref{step:log_law} and estimate (\ref{eq:friction_estimate})). 
\medskip

ii) As the thickness of the linear sub-layer is very small compared to other scales involved in the problem, a Taylor expansion allows to deduce from the no slip condition at the flow domain boundary a Navier BC at the computational walls ( (\ref{eq:navier_BC}) in \ref{step:calc_domain}). This is as if the linear sub-layer would exert a friction over the log sub-layer.  
\medskip

iii) The mesh yields natural numerical length scales $h_K$, where $h_K$ is a diameter of any $K \in {\cal T}_h$. Therefore, one takes $\nut$ of the form 
$\nut = h_K^2 | D \vv_h | $ on $K \in {\cal T}_h$ inside the computational domain, and $\nut = h_K u_\star (\vv, \x)$ on $K \in {\cal T}_h$  in the log layer (see \ref{step:eddy_viscosities}). 
\medskip

Once this modeling  process is completed, we get a model expressed in its variational form  over finite element space $\vbh\times \sph$, as described in    \ref{step:the_model}.  So far as we know, this model is totally new, and can be  generalized to more complex and realistic geometries thanks to a careful differential geometry analysis, which is a work under progress. 
\medskip

We consider all over the paper the steady-state case, which is in coherence with the fact that in a permanent regime and for a developed turbulence, mean fields are steady, which is not in contradiction with the fact that fluctuations might be time dependent. 
\medskip

We prove that this variational problem has a solution $(\vv_h, p_h ) \in \vbh\times \sph$ (Theorem \ref{le:stasmagoh}) which converges to a solution 
$(\vv, p)$ of the steady-state Navier Stokes Equation (NSE) with Navier BC (Theorem \ref{th:cvgsmagostat}). 
\medskip

The paper is organized as follows. We start with general setting. Then we derive from the NSE a description of the boundary layer, introducing the friction velocity. We specify the computational domain and Navier BC, and next we perform the finite element setting and get the model. Finally we state and prove Theorem \ref{le:stasmagoh} and Theorem \ref{th:cvgsmagostat}.

 \medskip

{\sl Acknowledgements:} R. Lewandowski thanks Professor Li Tatsien and the ISFMA at Fudan University, Shangha\"i, China, for their hospitality during spring 2012, where part of this work was performed. He also thanks E. Memin and the Fluminance team at INRIA Rennes for their support during the second term of the academic year 2012-2013. The research work of T. Chac\'on was partially supported by the Spanish Government Grant MTM2012-36124-C02-01.

\section{General framework}

\subsection{Channel flow} 

\step {\sl Geometry, equations and boundary conditions}. Let  $\Om_{f}$ be a channel periodic in the $x_1$ axis and $x_2$ axis, of height $1+2d$ in the $x_3$-axis, for a small parameter $d << 1$,  
\BEQ \Om_{f} =\left \{  \x=(x_1,x_2,x_3) \in \mathbb{T}_2 \times \R^3 \, \mbox{ s. t. }\,
 -d<x_3<1+d  \right \},  \EEQ
 where $  \mathbb{T}_2 $ is the two dimensional torus defined by
 $$ \mathbb{T}_2  = { \R^2 \over {\cal T}_2 } \qmbx{where}  {\cal T}_2 = {2 \pi \nbZ^2 \over L} ,$$
and $L>0$ is a given length scale. 
\medskip
Let $\Ga_f$ denote
\BEQ \Ga_f = \{\x\in \mathbb{T}_2 \times \R^3 \mbox{ s. t. }x_3 = -d \,\,\mbox{or} \,\, x_3 = 1+d \}. \EEQ
The steady-state Navier-Stokes equations with the no-slip boundary condition are as follows,
\begin{equation} \label{eq:ssNSE} \left \{
\begin{array}{rcll}
(\vv \cdot \g) \vv-\nu \Delta \vv+\nabla p&=&\fv & \qmbx{in} \Om_{f} , \\
\div \vv &=&0 & \qmbx{in} \Om_{f} ,\\
\vv &=& { \bf 0 } & \qmbx{on} \Ga_f.  \end{array}
\right .
\end{equation}
The source term ${\bf f}$ is a body force per mass unit, typically  the gravity.  Assuming ${\bf f}Ê\in L^2(\Om_{f} )^3 = {\bf L}^2 (\Om_f)$, we know that this equation has a solution $(\vv, p) \in W^{2, 3/2} (\Om)^3 \times W^{1, 3/2} (\Om)$ (see in \cite{temam6}), whose norms are bounded by constants that only depend on 
$\nu$, $||{ \bf f } ||_{0,2, \Om_{f} }$ and $d$, and $p$ is defined up to a constant. Uniqueness is known when 
$ ||{ \bf f } ||_{0,2, \Om_{f} } / \nu^2$ is small enough. 

\medskip

 \step {\sl Friction velocity.}\label{step:friction1} Let $(\vv, p)$ be any solution of \ref{eq:ssNSE}. We still denote by 
$\vv$ the trace of $\vv$ on $\Ga_f$. 
We deduce from trace Theorems and Sobolev Theorem that $\vv \in W^{1, 3} (\Ga_f )^3 = {\bf W}^{1, 3} (\Ga_f ) $. Therefore, it makes sense to consider $D \vv \cdot \nv$ on $\Ga_f$, where ${\bf n}$ denotes the outward-pointing unit 
normal vector at $\Gamma_f$, $D\vv = (1/2) (\g \vv + \g \vv^t)$.  We split the vector  $D \vv \cdot \nv$ into its tangential part and its normal part, 
\BEQ \label{eq:friction} D \vv \cdot \nv = (D \vv \cdot \nv)_{\tau} +  ( (D \vv \cdot \nv) \cdot \nv) \, \nv. \EEQ
Let  $v_\star \in L^6(\Ga_f)$ be defined on $\Ga_f$ by 
\BEQ \label{eq:esthache} v_\star = v_\star (\vv)(\x) = ( \nu | (D \vv \cdot \nv)_{\tau} (\x) | )^{1 \over 2}, \EEQ 
called the friction velocity associated to $\vv$ at $\x \in \Ga_f$.  

\subsection{Boundary layer description}

\step {\sl Length scale}. Condition of uniqueness to system (\ref{eq:ssNSE}) is not satisfied in a steady-state turbulent regime. Let 
${\cal S}$ be the set of solutions, which is a closed subset in ${\bf L}^2 (\Om_f)$. According  to \cite{chalew}, one can construct a probability measure $\mu$ on ${\cal S}$. We consider the following velocity friction $w_\star \in {\bf L}^6(\Ga_f)$ defined by,
\BEQ  w_\star = \int_{ {\cal S} } v_\star (\vv) d\mu( \vv). \EEQ
We finally define the meanfriction velocity by 
\BEQ u_\star = {1 \over L} || w_\star ||_{0, 2, \Ga_f} \in \R \EEQ 
to which is associated the typical length scale $\lambda$ that characterises the boundary layer,
\BEQ \label{eq:echelle}  \lambda = {\nu \over u_\star } , \EEQ
assuming  $  u_\star \not= 0$. 
  \medskip
  
{\sl We conjecture that $u_\star \rightarrow \infty$ when $|| {\bf f} ||_{0,2, \Om_f} \rightarrow  \infty$. }
  \medskip 
  
 \step {\sl Main assumption about the boundary layer structure.} \label{step:BLass}  We focus on the bottom of $\Om_f$, $\{ x_3 = -d \}$, assuming that the boundary layer at the top $\{ x_3 = 1+d \}$ has a similar structure.
According to experiments (see in \cite{schli}), we assume that in the boundary layer, the mean fluid velocity has a constant direction and only depends on the variable $x_3$, which means 
${\bf \vv}(\x) =  v(x_3) {\bf e}$, for some fixed unit vector ${\bf e}$. Without  loss of generality, we can assume that ${\bf e} = {\bf e}_1$ is the unit vector pointing along the $x_1$-axis. 

\medskip
Notice that any plane $P$ of the form  
$P = \{ x_3 = h \}$ included in the boundary layer, and any vector ${\bf N}$ orthogonal at $P$  being given, our assumption yields in particular 
$\vv \cdot {\bf N} = 0$ at $P$. 
\medskip

 \step {\sl Log law.} \label{step:log_law}  Experiments and suitable assumptions about turbulence \cite{chalew, schli} indicate that the boundary layer 
can be decomposed into two sub layers:  
\begin{itemize}
\item near the boundary where the velocity profile $v$ is linear  (linear sub layer), 
\item the next  sub-layer  specified by a log profile (log layer). 
\end{itemize}
To be more specific, 
we introduce the dimensionless variable 
\BEQ z^+ = {x_3 \over \lambda} ,\EEQ
and we consider the following continuous function defined on $[0, z^+_{max} ]$ by 
\BEQ \label{eq:wallaw}
L(z^+) = 	\left \{ \begin{array}{ccl}
z^+ & \mbox{if}  &0 \le z^+ \le z_0^+ \\
\disp \frac{1}{\kappa} \, \log\left (\frac{z^+}{z_0^+}\right )+z_0^+ & \mbox{if}  &  z_0^+ \le z^+ \le z^+_{max}  ,
 \end{array}
 \right .
\EEQ
where $\kappa \approx 0,41$ is the Von K\'arm\'an constant. In practical calculations, one takes $z_0^+ \approx 20$, and $z^+_{max} \approx 100$, that measures the thickness of the logarithmic boundary layer, taken to be equal to $100 \lambda$. According to experiments   \cite{schli}, boundary layer thickness goes to zero as the Reynolds number goes to infinity. 
\medskip

The profile $v$ in the boundary layer at the bottom of $\Om_f$  is given by the formula 
\BEQ \label{eq:velocity_log_law} v(x_3) = u_\star L \left ( x_3 \over \lambda \right ). \EEQ
A similar description applies to the boundary layer at the top of $\Om_f$, $\{z= 1+d \}$. 
\medskip

 \step {\sl Friction velocity expressed as a function of the velocity.} \label{step:friction}  We still focus on the bottom. Any $x_3 >0$ being given, Let 
\BEQ \label{eq:fonction_F} F (\beta) = \beta L (\alpha \beta), \quad \alpha = {x_3 \over \nu}. \EEQ
With this notation, equation (\ref{eq:velocity_log_law}) may be written as 
\BEQ \label{eq:friction_velocity} 
v = F (u_\star), \EEQ
thanks to definition (\ref{eq:echelle}). 

\begin{lemma} \label{lem:invertible} Let $F : [0,+\infty) \rightarrow [0,+\infty)$ be defined by by (\ref{eq:fonction_F}). The function $F$ is invertible, so that equation (\ref{eq:friction_velocity}) can be written as $u_\star = F^{-1} (v)$ at each given $x_3$. 
\end{lemma}

{\bf Proof}. 
We observe that the function $L$ satisfies
\BEQ\label{eq:wl1}
\disp \lim_{x \to 0^+} \frac{L(x)}{ x}=C_1,
\EEQ
\BEQ\label{eq:wl2}
\lim_{x \to \infty} \frac{L(x)}{\log x}=C_2,
\EEQ
where $C_1$ and $C_2$ are non-zero constants. As $L$ is strictly increasing and continuous in $(0,+\infty)$, then $F$ is strictly increasing and con\-ti\-nuous in $(0,+\infty)$. Also, by (\ref{eq:wl1}) $F$ is continuous at $\beta=0$ with $F(0)=0$. Moreover, by (\ref{eq:wl2}) 
$\disp \lim_{x \to \infty} F(x)=+\infty$. Then $F$ is bijective from $[0,+\infty)$ onto $[0,+\infty)$, which yields the invertibility of $F$ as claimed. 
\qed
\medskip

\begin{lemma} 
Denote $h = F^{-1}$. Then there exist a constant $C = C(x_3) >0$, bounded, such that 
\BEQ \label{eq:fonction_h} \forall \, \gamma >0, \quad  h(\gamma) \le C (x_3) \, (1+\gamma) \le C (1+\gamma) , \quad C = \sup C (x_3). \EEQ
\end{lemma}

{\bf Proof}.  Then $h : [0,+\infty) \mapsto [0,+\infty)$ is bijective and continuous. Also,
$$
\lim_{\gamma \to \infty}\frac{{h}(\gamma)}{\gamma}=\lim_{t \to \infty}\frac{t}{F(t)}=\lim_{t \to \infty}\frac{1}{L(\alpha t)}=\lim_{t \to \infty}\frac{1}{\log(\alpha t)}\frac{\log (\alpha t)}{L(\alpha t)}=0.
$$
The conclusion is a consequence of the continuity of ${h}$. \qed
\medskip

We deduce from Lemma \ref{lem:invertible}, inequality (\ref{eq:fonction_h}) and because top and bottom layers have the same structure,  that the friction velocity can be calculated at each $\x \in BL$ from the velocity $\vv$, and satisfies the estimate
\BEQ \label{eq:friction_estimate}Ê0 < u_\star = u_\star ( \vv, \x) \le C (1 + | \vv | ). \EEQ

\section{Turbulence model} 

\subsection{Geometry and meshing}

\step{\sl Calculation domain}. \label{step:calc_domain}
From now we assume that the boundary layer is included in the union of two strips, 
\BEQ \label{eq:boundary_layer} BL =  \{ -d \le x_3 \le D/2 - d \} \cup \{ 1+ d - D/2 \le x_3 \le 1+d \}, \EEQ
 where 
$ d < D << 1$, $d$ being the order of the linear sub layer, $D$ the thikness of the global boundary layer. Standard numerical simulations are carried out in a sub domain of the flow 
domain that does not include the boundary layer at all, using a wall law  \cite{chalew, MP94, pares} at artificial boundaries (walls). Our model 
includes  the log layer, using a Navier BC based on a Taylor expansion as shown below. 

\medskip

The computational domain is 
 \BEQ \Om =\left \{  \x=(x_1,x_2,x_3) \in \mathbb{T}_2 \times \R^3 \,\,\mbox{s. t.}\,\,
0<x_3<1 \right \},  \EEQ
the artificial wall being defined by 
\BEQ \Ga_{w} = \{ \x\in \mathbb{T}_2 \times \R^3 \,\,\mbox{s. t.}\,\,x_3 = 0  \,\,\mbox{or}\,\, x_3 = 1 \}. \EEQ

 \step{\sl Boundary conditions}. As in above, we focus on the bottom layer. By a Taylor expension we get 
\BEQ \label{eq:DL1}  0 = v \vert_{x_3 = -d}  \approx    v \vert_{x_3 = 0}      - d {\p v \over \p x_3}\vert_{x_3 = 0} .         \EEQ
From the view point of the domain $\Om$, $v = \vv_\tau \vert_{\Ga_w}$, and $\p / \p x_3 = - \p / \p \nv$ at $\Ga_w$, where 
$\vv_\tau$ is the tangential part of $\vv$, defined by 
\BEQ \label{eq:vtau} \vv = \vv_\tau + (\vv \cdot \nv) \nv, \EEQ 
by still denoting $\vv$ the trace of $\vv$ at $\Ga_w$, so far no risk of confusion occurs. Therefore, by remarks in \ref{step:BLass}
together with (\ref{eq:DL1}), we get  
\BEQ   \label{eq:navier_BC} \vv \cdot \nv \vert_{\Ga_{w} } = 0, \quad { \p \vv_\tau  \over \p \nv}Ê\vert_{\Ga_{w} }  = - {1 \over d} \vv_\tau, \EEQ
which is a Navier boundary condition at the artificial wall, that expresses in some sense that the linear sub-layer exerts a friction 
on the log layer. Hence, system (\ref{eq:ssNSE}) becomes in $\Om$,
\begin{equation} \label{eq:ssNSE_comp} \left \{
\begin{array}{rcll}
(\vv \cdot \g) \vv-\nu \Delta \vv+\nabla p&=&\fv & \qmbx{in} \Om , \\
\div \vv &=&0 & \qmbx{in} \Om ,\\
\vv \cdot \nv &=& { \bf 0 } & \qmbx{on} \Ga_w, \\
\displaystyle  - { \p \vv_\tau  \over \p \nv}Ê&=& \displaystyle     {1 \over d} \vv_\tau & \qmbx{on} \Ga_w.
  \end{array}
\right .
\end{equation}
Navier-Stokes equations with Navier boundary conditions was studied before \cite{HDB05, HDB06, Berselli3, BMR, verf}, and existence 
of a solution to system (\ref{eq:ssNSE_comp})  is already ensured.   

 \medskip
  \step{\sl Variational formulation.}
 Let us define the spaces
$$
\vb=\{ \wv \in {\bf H}^1(\Om), \quad \wv \cdot \nv \vert_{\Ga_w} = 0 \},
$$$$
\lpp=\{ q \in L^2(\Om), \quad \, \int_{\Omega_c} q \, d \x=0\},
$$
by reminding that ${\bf H}^1(\Om) = H^1(\Om)^3$. 
Strictly speaking, The space $\lpp$ is isomorphic to the quotient space $L^2 (\Om) / \R$, endowed with the usual quotient norm 
\BEQ ||Ê\stackrel{.} p ||_M =  \inf _{p \in Ê\stackrel{.} p} || p ||_{0, 2, \Om} .\EEQ
It also may be viewed as a closed subspace of $L^2(\Om)$ endowed with the $L^2(\Om)$ norm. 
\medskip

The space $\vb$ is endowed with the ${\bf H}^1$ norm, denoted $ || \cdot ||_{1, 2, \Om}$.  As a consequence of Korn's inequality, the following usefull estimate holds, 
\BEQ \label{eq:norms}  \forall \, \vv \in \vb, \quad || \vv ||_{1,2, \Om} \le C ( || D \vv ||_{0,2, \Om} + || \vv ||_{0, 2, \Ga_w}) ,\EEQ  
of proof of which being carried out in \cite{BMR}. 
\medskip

Let  $a$, $b$ and $G$ the forms defined by
\begin{eqnarray} \label{eq:formb}
a(\vv,\wv)&=&\nu\,(D \vv,D \wv)_{\Om}, \label{eq:forma0}\\
b(\zv;\vv,\wv)&=&{1 \over 2} \left[  ( ( \zv \cdot \nabla)\,   \vv, \wv)_{\Om} - ( (\zv \cdot \nabla)\,  \wv, \vv)_{\Om}\right ] \\
  G(\vv,\wv )& =&\displaystyle\frac{\nu}{d}(\vv_\tau,\wv_\tau)_{\Gamma_{w}},
\end{eqnarray}
for $\zv,\, \vv,\,\wv \in {\bf H}^1(\Om)$. Recall that when $\zv,\, \vv,\,\wv  \in \vb$ and $\div \zv = 0$, then 
$b(\zv; \vv, \wv) = ( (\zv \cdot \nabla) \, \vv, \wv)_\Om$, and $(\g \zv, \g \wv)_\Om =  (D \zv, D \wv)_\Om$. Also remark that when $\vv \in \bv$, then $\vv= \vv_\tau$ at $\Ga_w$.
\medskip

We say that a pair $(\vv,p)\in \vb \times  \lpp$ is a weak solution of the boundary value (\ref{eq:ssNSE_comp}) if it satisfies
\BEQ  \label{eq:nsh} \left \{ \begin{array}{rcl}
b(\vv;\vv,\wv)+a(\vv,\wv)-(p,\div\wv)_{\Om} +  G(\vv,\wv  ) &=& \langle \fv, \wv \rangle,\\
(\div\vv,q)_{\Om} &=& 0,
\end{array} \right .
\EEQ
for any $(\wv,q) \in \vb \times \lpp$.
\medskip

\step {\sl A priori estimate and existence result}. Assume ${\bf f} \in \vb'$. Let $(\vv, p)$ be any solution of Problem (\ref{eq:nsh}), and take $\vv = \wv$ in (\ref{eq:nsh}). From the standard formula $b(\vv;\vv,\vv) = 0$ that holds since $\div \vv = 0$ and $\vv \cdot \nv = 0$ at $\Ga_w$, we get 
\BEQ \nu || D \vv ||_{0,2,\Om} + {\nu \over d} ||Ê\vv ||_{0, 2, \Ga_w} = \left < {\bf f}, \vv \right >, \EEQ
from where we deduce 
\BEQ || \vv ||_{1,2, \Om} \le C\kappa^{-1} || {\bf f} ||_{ \vb'}, \quad \kappa =  \min \left (  \nu, { \nu \over d} \right ),  \EEQ
by using  (\ref{eq:norms}).

\subsection{Finite element setting} 

\step{\sl Triangulation}. Let $D \subset \R^3$ denotes the sample box 
$ D = [0,L]^2 \times [0, 1]$. The computational domain $\Om$ may be viewed as the periodic reproduction of 
$D$ in the $x_1-x_2$ axes. Let $\{ {\cal T}_h \} _{(h>0)}$ be a regular familly of triangulation of $D$, compatible with the periodicity of the domain: The restriction of the grid to the planes $x_1=0$ and $x_1=D$ is the same, and the restriction of the grid to the planes $x_2=0$ and $x_2=D$ is the same. Reproducing this triangulation by periodicity, we get a regular triangulation of $\Om$, still denoted by  $\{ {\cal T}_h \} _{(h>0)}$. 

\medskip

  In the following, for  each $K \in {\cal T}_h$, $h_K = diam (K)$ denotes the diameter of $K$, and 
 \BEQ \displaystyle h = \max_{K \in {\cal T}_h}Êh_K. \EEQ

   \step {\sl Eddy viscosities}. \label{step:eddy_viscosities} We assume isotropy of the turbulence inside the domain defined by 
 $\Om_{in} = \Om \setminus BL$, the boundary layer $BL$ being defined by (\ref{eq:boundary_layer}). This yields to consider in $\Om_{in}$ the eddy viscosity $\nu_{t,in}$ to be of the same form as  in usual Sub-Grid-Models of Prandtl-Kolmogorov-Smagorinsky type, where following  \cite{chalew}, we take in each $K$  the length scale equal to $h_K$, leading to consider $\nu_{t,in}$ to be of the form
 \BEQ \label{eq:innvisc}
\nu_{t,in}(\vv)= C_s^2 \sum_{K\in \trh}  \, h_K^2 \, \mathbf{1}_K |D \vv|,
\EEQ
$C_s>0$ being an empirical constant, $\mathbf{1}_A$ denotes the characteristic function for any set $A$. 
\medskip

In the boundary layer part, $\Om_{w} = BL \cap \Om$, turbulence is no longer isotropic and depends on the friction velocity. Taking again $h_K$ as typical length scale and by a dimensional analysis argument \cite{chalew}, we define the eddy viscosity $\nu_{t,w}$ in $\Om_{w}$ by
\BEQ \label{eq:wallvisc}
\nu_{t,w}(\vv)= C_w \sum_{K\in \trh}  \, h_K \, \mathbf{1}_K u_\star(\vv, \x),
\EEQ
where $C_w>0$ is an empirical constant and $u_\star$ is expressed in \ref{step:friction}. 
\medskip

Finally, the eddy viscosity we consider is of the form 
\BEQ \nu_t = \nu_t (\vv) = \mathbf{1} _{\Om_{in} }  \nu_{t,in}(\vv) +   \mathbf{1} _{\Om_{w} }  \nu_{t,w}(\vv), \EEQ

   \step {\sl Finite element spaces}. The model is a mixed formulations, based upon pairs of finite element spaces $(\vbh,\sph) \subset \vb\times \lpp$, associated to the family of regular triangulations $\{\trh\}_{h>0}$ of $\Omega$ in the sense of Ciarlet \cite{ciarl}. We assume that the family of pairs of spaces $\{(\vbh,\sph)\}_{h>0}$ satisfies the following hypothesis:
\medskip

{\sc Hypothesis 1:} The family of spaces $\{\vbh \times \sph\}_{h>0}$ is an internal appro\-xi\-mation of $\vb\times \lpp$: For all $(\wv, p) \in \vb\times \lpp$ there exists a sequence $\{(\vhv,\ph)\}_{h>0}  $ such that $(\vhv,\ph) \in \vbh \times \sph$, and
$$
\lim_{h\to 0} (\nor{\vv-\vhv}{1,2,\Om}+\nor{p-\ph}{0, 2,\Om})=0.
$$
\par
{\sc Hypothesis 2:} The family of pairs of spaces $\{(\vbh,\sph)\}_{h>0}$ satisfies the uniform discrete inf-sup condition : There exists a constant $\alpha >0$ such that
\BEQ \label{eq:isdis}
\alpha \nor{q_h}{0, 2, \Om} \le \sup_{\whv \in \vbh} \frac{(\nabla \cdot \whv, q_h)_{\Om}}{\nor{\whv}{1,2,\Om}},\qmbx{for all} q_h \in \sph
\EEQ
There is a wide literature about finite element spaces satisfying those properties (Cf. \cite{bernraug},  \cite{brzfor}, \cite{gira}).

 \medskip
    \step {\sl The model}. \label{step:the_model} Our LES model is expressed by the following variational problem: 
 $$
\qmbx{Find}(\vhv,\ph) \in \vbh\times \sph\qmbx{such that for all} (\whv,\qh) \in \vbh\times \sph,
$$
\begin{equation} \label{eq:smagoh} \left \{
\begin{array}{rcl}
b(\vhv;\vhv,\whv)+a(\vhv,\whv)+c(\vhv;\whv) + \phantom {bbbbbbbb}  \\
G(\vhv,\whv  )-(\ph,\div\whv)_{\Om}&=&\langle \fv, \whv \rangle ,\\
(\div \vhv,\qh)_{\Om}&=&0 ;
\end{array}
\right .
\end{equation}
 the form $c$ being defined by
\BEQ \label{eq:forma1} \begin{array} {l}
c(\vv;\wv)=(\nut(\vv)\,D\vv,D \wv)_{\Om}\,\, \mbox{or} \\ c(\vv;\wv)=(\nu_{t,in}(\vv)\,D\vv,D \wv)_{\Om_{in}}+ (\nu_{t,w}(\vv)\,\partial_3\vv,\partial_3 \wv)_{\Om_{w}}. \end{array}
\EEQ
The second expression neglects the tangential eddy viscosity  in the boundary layer, which is very small compared to the normal one.
\section{Analysis of the model} 

\subsection{Technical results} 

We state in this sub-section some technical results concerning the eddy viscosities and the associated turbulent diffusion form $c$, that are needed by our analysis.
\medskip

    \step {\sl $L^\infty$ eddy viscosties estimates.}

\begin{lemma} \label{le:estnut}There exists a constant $C>0$ depending only on the aspect ratio of the family of triangulations $\{\trh\}_{h>0}$ such that
\BEQ \label{eq:estnut}
\nor{\nut(\vhv)}{0,\infty,\Omega} \le C\, h^{1/2}\,\nor{ \vhv}{1,2,\Omega},\qmbx{for all} \vhv \in \vbh.
\EEQ
\end{lemma}
 {\bf Proof.} We start with the internal part of the eddy viscosity $\nu_{t, in}$. 
Consider $\vhv \in \vbh$. As $\grad \vhv$ is piecewise continuous, there exists $K \in \trh$ such that
$$\nor{\nu_{t,in}(\vhv)}{0,\infty,\Omega}=\nor{\nu_{t,in}(\vhv)}{0,\infty,K}\le C_{S}^2\, h_{K}^2 \, \nor{\nabla \vhv}{0,\infty,K}.
$$
By a standard finite element inverse estimate (Cf. \cite{BMR04}), 
$$ \nor{\grad\vhv}{0,\infty,K} \le C\, h_{K}^{-3/2}\,  \nor{\grad\vhv}{0,2,K}$$ for some constant $C>0$ depending only on the aspect ratio of the family of triangulations. Then,
\BEQ
\label{eq:estnutin}
\nor{\nu_{t,in}(\vhv)}{0,\infty,\Omega} \le CC_{S}^2\, h_{K}^{2-3/2}\,\nor{\nabla \vhv}{0,2,K} \le CC_{S}^2\, h^{1/2}\,\nor{\nabla \vhv}{0,2,\Omega},
\EEQ

Next, we analyze the wall eddy diffusion $\nu_{t,w}$. There exists some element $K\in \trh$ such that
$$\nor{\nu_{t,w}(\vhv)}{0,\infty,\Omega}=\nor{\nu_{t,w}(\vhv)}{0,\infty,K}\le C_{w}\, h_{K}\,  \, (1+\nor{ \vhv}{0,\infty,K}),
$$
where in the last inequality we have used (\ref{eq:esthache}). Using the inverse estimate (Cf. \cite{BMR04}), $ \nor{\vhv}{0,\infty,K} \le C\, h_{K}^{-1/2}\,  \nor{\grad\vhv}{0,2,K}$ we deduce
$$\nor{\nu_{t,w}(\vhv)}{0,\infty,\Omega} \le  C'\,C_{w}\, h^{1/2}\,\nor{\nabla \vhv}{0,2,\Omega},\qmbx{for some constant} C'>0.
$$
Combining this estimate with (\ref{eq:estnutin}) and $\nor{\nabla \vhv}{0,2,\Omega} \le \nor{ \vhv}{1,2,\Om}$, (\ref{eq:estnut}) follows. \qed
\medskip

    \step {\sl Turbulent diffusion operator properties.}
  
\begin{lemma}\label{le:estopera}
The form $c$ defined by (\ref{eq:forma1}) satisfies the following properties:
\begin{enumerate}
\item[i)] $c$ is non-negative, in the sense that
$$
c(\vv;\vv) \ge 0,\qmbx{for all} \vv \in \sh{1}{\Om}^3.
$$
\item[ii)] Assume that the family of triangulations $\{\trh\}_{h>0}$ is regular. Then, for any $\vhv,\,\whv \in \vbh$,
\begin{eqnarray}\label{eq:estop}
|c(\vhv;\whv)| \le C\, h^{1/2}\,\,\nor{ \vhv}{1,2,\Omega}^2 \nor{ \whv}{1,2,\Omega},
\end{eqnarray}
for some constant $C>0$ depending only on $d$, $\Omega$ and the aspect ratio of the family of triangulations.
\item[iii)] Assume that the family of triangulations $\{\trh\}_{h>0}$ is regular. Let $\{\vhv\}_{h>0}$ and $\{\whv\}_{h>0} $ be two sequences such that $\vhv,\,\whv \in \vbh$. Then, if both sequences are bounded in $\shb{1}{\Om}^{d}$,
\BEQ \label{eq:a1tocero}
\lim_{h \to 0} c(\vhv;\whv)=0
\EEQ
\end{enumerate}
\end{lemma}
 {\bf Proof.} 
\begin{enumerate}
\item[i)] Let $\vv \in \shb{1}{\Om}$. Then,
$$
c(\vv;\vv) = \int_{\Om} \nut(\vv)\, | D \vv|^{2}\, d\x \ge 0.
$$
\item[ii)] By estimate (\ref{eq:estnut}),
\begin{eqnarray*}
|c(\vhv;\whv)|&\le&\nor{\nut(\vhv)}{0,\infty,\Omega} \,  \nor{\vhv}{1,2,\Omega}\nor{\whv}{1,2,\Omega} \\
&\le& C\, h^{1/2}\,\nor{ \vhv}{1,2,\Omega}^2\nor{ \whv}{1,2,\Omega}.
\end{eqnarray*}
\item[iii)]
Statement (\ref{eq:a1tocero}) directly follows from (\ref{eq:estop}).
\end{enumerate}
\qed

\subsection{Existence result} 

Problem (\ref{eq:smagoh}) is a set of non-linear equations in finite dimension. These non-linearities are due to several effects: the convection operator, the eddy viscosity, and the wall-law boundary conditions. The space $\vb$ is a closed sub-space of ${\bf H}^1 (\Om)$. 
Our main result is the following. 
\begin{theorem} \label{le:stasmagoh}
 Let $\{\trh\}_{h>0}$ be  a regular family of triangulations of the domain $\Omega$. Let $\{(\vbh,\sph)\}_{h>0}$ be a family of pairs of finite element spaces satisfying Hypotheses 1 and 2. Then for any  $\fv \in \vb'$ the variational problem (\ref{eq:smagoh}) admits at least a solution, that satisfies the estimates
\begin{eqnarray}
\nor{ \vhv}{1,2,\Om}&\le& C \kappa^{-1} \nor{\fv}{\vb'}, \quad \kappa = \min \left (\nu,  {\nu \over d} \right ) \label{eq:estsmavel} \\
\nor{ \ph}{0,2,\Om}&\le& C  \kappa^{-1} || {\bf f} ||_{\vb'} \left ( \kappa^{-1}  || {\bf f} ||_{\vb'} [1 + h^{1/2}] +  \nu + {1 \over d} +1 \right )
.\label{eq:estsmapres}
\end{eqnarray}
where $C >0$ is a constant depending only on $d$, $\Omega$ and the aspect ratio of the family of triangulations.
\end{theorem}
\medskip

 {\bf Proof.}. We prove the existence of solution in two steps.
\medskip

 \step {\sl  Step 1: Existence of the velocity.} Let us define the mapping $\Phi_h:\vbh \to \vbh'$ as follows: Given $\zhv \in \vbh$,
$$
\begin{array}{rcl}
\langle \Phi_h(\zhv),\whv\rangle&=& b(\zhv;\zhv,\whv)+a(\zhv,\whv)+c(\zhv;\whv) + G(\zhv,\whv  )-\langle \fv, \whv \rangle ,
\end{array}
$$
for any $\whv \in \vbh$. This equation has a unique solution as its r.h.s. defines a linear bounded functional on $\vbh$.  Moreover, the functional $\Phi_h$ is continuous as all functions that appear in its definition are continuous on the finite-dimensional space $\vbh$.
\par
 Consider the sub-space $Z_{h}$ of $\vbh$ defined by
$$
Z_{h}=\{ \whv \in \vbh \qmbx{such that} (\div \whv, \qh)=0,\qmbx{for all} \qh \in M_h\, \}.
$$
$Z_{h}$  is a non-empty closed sub-space of $\shb{1}{\Omega}$. Then it is a Hilbert space endowed with the $\shb{1}{\Omega}$ norm. Let $\zhv \in Z_{h}$. Then, as $b(\zhv;\zhv,\zhv) = 0$ and $c$ is non-negative,
\begin{eqnarray*}
\langle \Phi_h(\zhv),\zhv\rangle&\ge& 
a(\zhv,\zhv) + G (\zhv, \zhv)- \langle \fv,\zhv \rangle   \\
&\ge& \nu  \,\nor{D(\zhv)}{0,2,\Om}^{2} + {\nu \over d}Ê\nor{\zhv}{0,2, \Ga_w}^2 -\nor{\fv}{\vb'} \nor{ \zhv}{1,2, \Om} \\
&\ge&  {C \kappa \over 2} \nor{\zhv}{1,2, \Om}^2 - { \nor{\fv}{\vb'} ^2 \over 2 C \kappa } 
\end{eqnarray*}
where we have used (\ref{eq:norms}) and Young's inequality.
We deduce 
\BEQ\forall \, \zhv \in Z_{h} \qmbx{such that} \nor {\zhv}{1,2, \Om} = {\nor {\fv}{ \vb'} \over C \kappa} , \quad 
\langle\Phi_h(\zhv),\zhv\rangle_{\sh{1}{\Omega}}\ge 0,Ê\EEQ 
Consequently, by a classical variant of Brouwer's Fixed Point Theorem (Cf. \cite{temam6}), the equation
\BEQ \label{eq:eqszh}
   b(\vhv;\vhv,\whv)+a(\vhv,\whv)+c(\vhv;\whv) + G(\vhv,\whv  )=\langle \fv, \whv \rangle \quad \forall \,  \whv   \in Z_{h}
  \EEQ
 admits a solution $\vhv \in Z_{h}$ such that $\displaystyle \nor{\vhv}{1,2,\Om} \le {\nor {\fv}{ \vb'} \over C \kappa}$, which precisely is  (\ref{eq:estsmavel}) by changing $C$ in $C^{-1}$. 
\medskip

 \step {\sl Step 2: Existence of the pressure.} Let the operator ${\cal G}_h:M_h \mapsto \vbh'$ defined by
 $$
 \forall q_h \in M_h,\, \langle {\cal G}_h(q_h),\vhv \rangle =(\div \vhv,q_h)_\Omega, \qmbx{for all} \vhv \in \vbh.
 $$ 
 Then $Z_h= Im({\cal G}_h)^\perp$. As $Im({\cal G}_h)$ is closed, then $Z_h^\perp= Im({\cal G}_h)$. As $\vhv$ is a solution of (\ref{eq:eqszh}), then $\Phi_h(\vhv) \in Z_{h}^{\perp}$. Consequently, there exists some discrete pressure $\ph$ such that $\langle \Phi_h(\vhv), \whv \rangle = (\div \vhv,\ph)_\Omega$, for all $\whv \in \vbh$. Thus, the pair $(\vhv,\ph)$ solves problem (\ref{eq:smagoh}). The estimate for the norm of the pressure is obtained via the discrete inf-sup condition (\ref{eq:isdis}),
 $$
 \nor{\ph}{0,2,\Omega} \le \alpha^{-1}\, \nor{\Phi_h}{\vbh'} ,
 $$
 for some constant $\alpha >0$. By estimates (\ref{eq:estop}) and some standard estimates,
\begin{eqnarray*}
\langle \Phi_h(\vhv), \whv\rangle &\le& C\, \left [ \nor {\vhv}{1,2,\Om}^2 (1+ C h^{1/2} ) + \nu \nor {\vhv}{1,2,\Om} \left (1 + {C \over d} \right ) \right ] \nor{\whv}{1,2,\Om} \\ 
&+&\nor{\fv}{\vb'}\,\nor{\whv}{1,2,\Om}.
\end{eqnarray*}
Then, the pressure estimate (\ref{eq:estsmapres}) follows from the velocity estimate (\ref{eq:estsmavel}). \qed

\subsection{Convergence} 

We now prove the convergence of the solution provided by method (\ref{eq:smagoh}) to a weak solution of the Navier-Stokes boundary value problem model (\ref{eq:ssNSE}).
\begin{theorem} \label{th:cvgsmagostat}
 Under the hypotheses of Theorem \ref{le:stasmagoh}, the sequence $\{(\vhv,\ph)\}_{h>0}$ contains a sub-sequence strongly convergent in $\shb{1}{\Omega}^2 \times \lp{2}{\Om}$ to a weak solution $(\vv,p) \in  \vb \times \lpz{2}{\Om}$ of the steady Navier-Stokes equation (\ref{eq:ssNSE}). If this solution is unique, then the whole sequence converges to it.
\end{theorem}

 {\bf Proof.}  The proof is divided into 7 steps. 
\medskip

 \step{\sl Extracting sub sequences.}  By estimates (\ref{eq:estsmavel}) and (\ref{eq:estsmapres}), the sequence $\{(\vhv,\ph)\}_{h>0}$ is bounded in the space $\vb \times \lpz{2}{\Om}$ which is is a Hilbert space. Therefore, this sequence contains a subsequence, that we denote in the same way, weakly convergent in $\vb \times \lpz{2}{\Om}$ to some pair $(\vv,p)$. As the injection of $\sh{1}{\Om}$ in $\lp{q}{\Om}$ is compact for $1\le q <6$, we may assume that the subsequence is strongly convergent in $\lpb{q}{\Om}$ for $1\le q <6$, and so in particular in $\lpb{4}{\Om}$. 
\medskip

 Also, the injection of $\shb{{1/2}}{\Ga_w}$ into $\lp{2}{\Ga_w}$ is compact. Then we may assume that the sequence $\{{\bf v}_{h_{{|_{\Ga_w}}}}\}_{h>0}$ is strongly convergent to ${\bf v}_{{|_{\Ga_w}}}$ in $\lpb{2}{\Ga_w}$.

\medskip

 \step{\sl Taking the limit in the diffusion terms.} Let $(\wv,q) \in \vb \times \lpz{2}{\Om}$. By Hypothesis 1, there exists a sequence $\{(\whv,\qh)\}_{h>0}$ such that $(\whv,\qh) \in \vbh \times \sph$ which is strongly convergent in $\shb{1}{\Omega} \times \lp{2}{\Om}$ to $(\wv,q)$.
\medskip

 As $a$ is bilinear and continuous, 
\BEQ \disp \lim_{h \to 0} a(\vhv,\whv)=a(\vv,\wv).
\EEQ  
Next, since the sequences $\{\vhv\}_{h>0}$ and $\{\whv\}_{h>0}$ are bounded in $\shb{1}{\Om}$, we deduce from Lemma \ref{le:estopera},
\BEQ \disp
\lim_{h \to 0} c(\vhv;\whv)=0. \EEQ
Finally it is straightforward to check that 
\BEQ  \disp
\lim_{h \to 0} G(\vhv;\whv)=0. \EEQ
 \step{\sl Taking the limit in limit in the convective term.} We have 
\begin{eqnarray*}
|(\vhv\cdot \grad\vhv,\whv)_{\Om}-(\vv\cdot \grad\vv,\wv)_{\Om}|&\le \\
|((\vhv-\vv)\cdot \grad\vhv,\whv)_{\Om}|+|(\vv\cdot \grad(\vhv-\vv),\wv)_{\Om}|
+|(\vv\cdot \grad\vhv,\whv-\wv)_{\Om}| &\le \\
 \nor{\vhv-\vv}{0,4,\Om} \,\nor{\grad \vhv}{0,2,\Om}\,  \nor{\whv}{0,4,\Om}  \\
+ \sum_{i,j=1}^{3}|(\partial_{j}(v_{hi}-v_{i}),v_{j}w_{i})_{\Om}|+\nor{\vv}{0,4,\Om} \,\nor{\grad \vhv}{0,2,\Om}\,  \nor{\whv-\wv}{0,4,\Om},
\end{eqnarray*}
where we denote $\vhv=(v_{h1},v_{h2}, v_{h3})$. All terms in the r.h.s. of the last inequality vanish in the limit because $\{\vhv\}_{h>0}$ is strongly convergent in $\lpb{4}{\Om}$, $\{\partial_{i}v_{hi}\}_{h>0}$ is weakly convergent in $\lp{2}{\Om}$ and $\{\whv\}_{h>0}$ is strongly convergent in $\shb{1}{\Om}$. Then,
\BEQ
\disp
\lim_{h \to 0} ( (\vhv\cdot\,  \grad\vhv),\whv)_{\Om}=( (\vv \g\,  \vv),\wv)_{\Om}.
\EEQ
Similarly, $
\disp
\lim_{h \to 0} ( (\vhv \cdot \g)\,   \whv,\vhv)_{\Om}=((\vv\cdot \grad\wv),\vv)_{\Om},
$
and then
$$
\lim_{h \to 0} b(\vhv; \vhv,\whv)=b(\vv; \vv,\wv).
$$

 \step{\sl Taking the limit in the pressure terms.}
Since $\{\div \vhv\}_{h>0}$ is weakly convergent in $\lp{2}{\Om}$ to $\div \vhv$ and $\{\qh\}_{h>0}$ is strongly convergent in $\lp{2}{\Om}$ to $q$,
$$
\lim_{h \to 0} (\div \vhv,\qh)_{\Om}=(\div \vv,q)_{\Om}.
$$
Finally, we obviously have 
$$ \lim_{h \to 0} (\ph,\div \whv )_{\Om}=(p ,\div \wv)_{\Om}.$$
Consequently, the pair $(\vv,q)$ is a weak solution of Navier-Stokes equations (\ref{eq:nsh}). 
\medskip

 \step{\sl Strong convergence of the velocities}. Set $\whv=\vhv$ in (\ref{eq:smagoh}). Then
$$
\nu \,\nor{D \vhv}{0,2,\Om}^{2} + {\nu \over d}  \nor{\vhv} {0,2, \Ga_w} =\langle \fv,\vhv \rangle - c(\vhv;\vhv). 
$$
By Lemma \ref{le:estopera} iii), $\disp \lim_{h\to 0}  c(\vhv;\vhv)=0$. Therefore, 
$$
\lim_{h\to 0} \left ( \nu \,\nor{D \vhv}{0,2,\Om}^{2} + {\nu \over d}  \nor{\vhv} {0,2, \Ga_w} \right ) =
 \langle \fv,\vv \rangle =\nu \,\nor{D \vv}{0,2,\Om}^{2} + {\nu \over d}  \nor{\vv} {0,2, \Ga_w}^2,
$$
where the last equality occurs because $(\vv,q)$ is a weak solution of Navier-Stokes equations (\ref{eq:nsh}). As $\vb$ is a Hilbert space and $\{\vhv\}_{h>0}$ is weakly convergent to $\vv$, this proves the strong convergence, since 
\BEQ \wv \rightarrow \left ( \nu \,\nor{D \wv}{0,2,\Om}^{2} + {\nu \over d}  \nor{\wv} {0,2, \Ga_w}^2\right )^{1 \over 2} \EEQ
is a norm equivalent to the ${\bf H}^1 (\Om)$ norm by (\ref{eq:norms}). 
\medskip

 \step{\sl Strong convergence of the pressures}. We use the discrete inf-sup condition to estimate $\nor{\ph-p}{0,2,\Om}$. There exists a sequence $\{P_{h}\}_{h>0}$ such that $P_{h}\in \sph$ for all $h>0$ which is strongly convergent in $ \lpz{2}{\Om}$ to $p$. We shall show that $\disp \lim_{h \to 0}\nor{\ph-P_{h}}{0,2,\Om} =0$. 
 Let $\whv \in \vbh$. We have
\begin{eqnarray*}
 (\ph-P_{h},\div \whv)&=& b(\vhv;\vhv,\whv)-b(\vv;\vv,\whv) + a(\vhv-\vv,\whv)+c(\vhv;\whv)\\
 &+&  G(\vhv-\vv,\whv  ) + (p-P_{h},\div \whv).
 \end{eqnarray*}
 As
 \begin{eqnarray*}
 b(\vhv;\vhv,\whv)-b(\vv;\vv,\whv) &=& b(\vhv;\vhv-\vv,\whv)+b(\vhv-\vv;\vv,\whv) \\
 &\le& C\, \nor{\vhv-\vv}{1,2,\Om} \, (\nor{\vhv}{1,2,\Om}+\nor{ \vv}{1,2,\Om}),
  \end{eqnarray*}
 using (\ref{eq:estop}) and the continuity of $a$ we deduce
\begin{eqnarray*}
 &&(\ph-P_{h},\div \whv)\le   C\, \left [ \nor{\vhv-\vv}{1,2,\Om} \, (\nor{\vhv}{1,2,\Om}+\nor{ \vv}{1,2,\Om}) + \nu \nor{D (\vhv-\vv)}{0,2, \Om} \right. 
 \\
&+&h^{1/2}\, \nor{ \vhv}{1,2,\Om}^{2} + \left .\disp \frac{\nu}{d}\,\nor{\vhv- \vv}{0,2, \Ga_w} + \nor{p-P_{h}}{0,2,\Om} \,\right ] \nor{\whv}{1,2,\Om}.
 \end{eqnarray*}
 As $\disp \lim_{h \to 0}\nor{\vhv- \vv}{1,2, \Om}=0$, then by Hypothesis 2, $\disp \lim_{h \to 0}\nor{\ph-P_{h}}{0,2,\Om} =0$. Then $\ph$ strongly converges to $p$ in $\lp{2}{\Om}$.
 \medskip
 
 \step{\sl Uniqueness.}  It remains to prove that if the Navier-Stokes equations (\ref{eq:ssNSE}) admit a unique solution $(\vv,p)$, then the whole sequence $\{(\vhv,p)\}_{h>0}$converges to it. This is a standard result that holds when compactness arguments are used, which is proved by reductio ad absurdum: Assume that the whole sequence does not converge to $(\vv,\ph)$. Then there exists a sub-sequence of $\{(\vhv,\ph)\}_{h>0}$ that lies outside some ball of $\vb \times \lpz{2}{\Om}$ with center $(\vv,p)$. Then the preceding compactness argument proves that a sub-sequence of this sub-sequence would converge to the unique solution $(\vv,p)$, what is absurd.
\qed


\end{document}